# Challenges of Designing and Developing Tangible Interfaces for Mental Well-being

**Kieran Woodward**
Nottingham Trent University
Nottingham, UK
Kieran.woodward@ntu.ac.uk

**Eiman Kanjo**
Nottingham Trent University
Nottingham, UK
Eiman.kanjo@ntu.ac.uk

**David Brown**
Nottingham Trent University
Nottingham, UK
David.brown@ntu.ac.uk

## ABSTRACT

Mental well-being technologies possess many qualities that give them the potential to help people receive assessment and treatment who may otherwise not receive help due to fear of stigma or lack of resources. The combination of advances in sensors, microcontrollers and machine learning is leading to the emergence of dedicated tangible interfaces to monitor and promote positive mental well-being. However, there are key technical, ergonomic and aesthetic challenges to be overcome in order to make these interfaces effective and respond to users' needs. In this paper, the barriers to develop mental well-being tangible interfaces are discussed by identifying and examining the recent technological challenges machine learning, sensors, microcontrollers and batteries create.

User-oriented challenges that face the development of mental well-being technologies are then considered ranging from user engagement during co-design and trials to ethical and privacy concerns.

**KEYWORDS**

Affective computing, Pervasive computing, Artificial intelligence, IoT, Tangible interfaces

**INTRODUCTION**

The capabilities and possibilities for tangible interfaces to have a positive impact on mental well-being are expanding rapidly. The increasing emergence of IoT and computationally-powerful devices is opening up new opportunities for delivering mental well-being monitoring in a more automated and accessible manner.

The World Health Organisation defines mobile health (mHealth) as "the use of mobile and wireless technologies to support the achievement of health objectives" [23]. Tangible interfaces go beyond mobile apps as they enable a person to interact with digital information through the physical environment. Mental well-being tangible interfaces possess many qualities including the ability to diagnose, promote positive mental well-being, collect and monitor well-being data remotely and transform well-being.

Tangible and embodied interaction improves accessibility, reduces stigma and reduces costs due to reduced needs for medical assistance; although there are many obstacles to overcome that need to address the usability and computational requirements of these systems. Addressing these challenges requires a combination of new design methodologies to cater to user needs and novel techniques to improve the functionality of these systems.

**BACKGROUND**

Technological solutions to aid mental well-being are rapidly increasing in popularity [12][13][2][1][11]. Numerous mental health apps have exceeded 10 million downloads including Calm harm and Headspace demonstrating the popularity of these technological solutions. More recently advancements in the IoT have led to the development of tangible interfaces for mental well-being [21][22]. Tangible interfaces have the potential to have a more significant impact than mobile apps as people are more likely to create stronger emotional attachments with physical devices rather than digital interfaces [15][17].

Previous mental well-being tangible interfaces have focused on promoting the communication and recording of mental well-being. EmoEcho [22] allowed two trusted partners to transmit their emotional well-being through haptic feedback in real-time and subtle stone [4] enabled students to privately transmit their emotions to a teacher through colours. Other tangible interfaces such as Emoball [6] and mood TUI [20] aimed to simplify the process of self-reporting emotions.

A range of tangible interfaces has previously been developed to sense mental well-being including wearables measuring heart rate variability (HRV) and skin conductance [9]. When the physiological sensors were paired with machine learning they could successfully classify stress with an accuracy of 97.4% after 5 minutes demonstrating the capabilities that could be incorporated into mental well-being tangible interfaces.

It is not only physiological sensors that can be used to sense and promote positive mental well-being. Force sensors have been used within a tangible ball that allows for the manipulation of music by sensing varying touch and motion patterns [5]. The research concluded that squeeze music could successfully be used for music therapy with children as it promoted positive emotions through tactile input and music.

Further devices have been devised to provide interventions; these devices use a variety of feedback mechanisms with the aim of improving mental well-being. Research such as Doppel [3] and good vibes [14] show tangible interfaces containing real-time interventional feedback can have positive impacts on mental well-being in stressful situations.

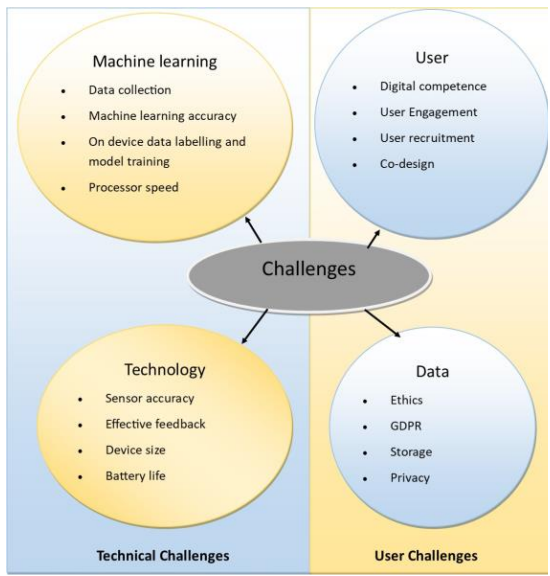

**Figure 1: Challenges associated with developing tangible Interfaces for mental wellbeing.**

Tangible interfaces for mental well-being have not yet utilised the full potential of recent advancements in IoT, sensors and artificial intelligence. This research aims to explore the common technological and user-oriented challenges faced when developing modern tangible and embodied interfaces for mental well-being. Figure 1 shows the different types of challenges associated with the design and development of tangible Interfaces for mental wellbeing.

**TECHNICAL CHALLENGES**

While advancements in CPUs, microcontrollers and sensors have resulted in smaller, more computationally powerful and accurate devices there are still many challenges to overcome when developing mental well-being tangible interfaces. One of the most significant issues currently faced is the physical size of the devices as users prioritise the look and feel of tangible interfaces [18] [20]. Mental well-being tangible interfaces must remain small and at the same time embed all of the required sensors, microcontroller and battery and while small powerful processors are available they increase the cost of such devices considerably.

**Sensor accuracy** remains a challenge as to infer mental well-being highly accurate sensors are vital. Recent advancements have resulted in off the shelf physiological sensors such as those measuring heart rate variability (HRV) and Electrodermal activity (EDA) to detect stress with similar accuracy as clinical grade sensors when they are paired with machine learning classifiers [16]. However, the machine learning classifiers must be first be trained with vast amounts of accurate real-world labelled data which can be difficult to obtain.

**Machine learning** provides clear benefits but requires computationally powerful devices to run. Ideally, the classification models would be run on-device as this would reduce latency and increase privacy as user data would never be transmitted off the device. New devices such as the Raspberry Pi zero are capable of running simple classification models on-device however their limited ram and larger size still create challenges. Alternatively, small microcontrollers such as the Arduino Nano could be utilised but they only allow for the collection of data. One option to maintain the small footprint of the Arduino Nano but still utilise machine learning is to combine it with Bluetooth Low Energy to transmit the data to a mobile application which can run the classification model.

Machine learning could also allow for the provision of personalised mental well-being inference. This provides a solution to one of the most challenging aspects when developing a one-size-fits-all tangible interface. Each device could be trained on the individual's sensory data resulting in the more accurate inference of mental well-being state. However, to achieve personalised inference a large amount of data is required from each individual when experiencing different states of well-being. A novel solution to this is the development of tangible interfaces containing the necessary sensors in addition to embedded techniques to self-label allowing for the collection of large amounts of real-world labelled data that can then be used to accurately train machine learning models.

**Battery life** is an increasing challenge as advancements in the development of more computationally powerful processors and additional sensors have outpaced advancements of batteries resulting in shortened battery life. Li-Po batteries are most suitable for mental well-being tangible interfaces as they are physically small and rechargeable, unlike 9v alkaline batteries which are commonly used with within the IoT. When developing tangible interfaces using Arduino Nanos and numerous physiological and environmental sensors Li-Po batteries between 500mAh and 1200mAh provided an average battery life of between 5 and 10 hours. While this is sufficient for the tangible interfaces if the machine learning classifiers were to run on-device this would reduce battery life significantly.

Many of the above challenges concern the detection of mental well-being but if tangible interfaces are to also improve well-being there are additional challenges to be overcome. The primary challenge is the use of different feedback mechanisms including auditory, visual and haptic feedback which will need to be examined to confirm their capability to reliably improve mental well-being for the general population.

**USER-ORIENTATED CHALLENGES**

In addition to the technical challenges that exist there are also many new user-oriented challenges that need to be addressed, many of which are unique to tangible interfaces for wellbeing.

Machine learning has been shown to provide many advantages when used to infer mental well-being but for it to be accurately trained a large amount of user data is first required making **user engagement** one of the largest challenges facing the accuracy of tangible interfaces. There are numerous reasons as to why user engagement is often low including digital competence as many users who experience mental health challenges may not possess high digital competence. However, research shows that tangible interfaces do not require high digital competence to be engaging which should result in any individual being able to use such devices [6][19].

**Stigma** is another issue which could reduce user engagement as traditionally there is much stigma around mental health and the diagnostic tools used. To combat this stigma the tangible interfaces should be designed to appear inconspicuous, one method to accomplish this is by embedding the sensors into pre-existing, familiar devices such as toys for children. By making the devices inconspicuous they are immediately more familiar and approachable for people to use and less stigmatised.

Given the stigma associated with mental health, **security** has to be a top concern for anyone developing or using tangible interfaces for mental well-being. Many users consider their mental health data to be highly personal and would not want it shared with any third parties. Because of the sensitive nature of the data on-device processing would be ideal as the data remains local and secure although this may not always be possible as additional computational power may be required. Legislation could help alleviate concerns as the General Data Protection Regulation (GDPR) [7] in the EU and EEA have attempted to give control to citizens over their personal data by ensuring they can access their data and understand how it is being processed. GDPR may help gain people's trust in mental well-being tangible interfaces as it allows people to assess how their personal data is stored and processed.

When developing devices to be used by people with mental well-being challenges it is essential to involve end users in the development process. Tangible interfaces for mental well-being were explored at focus groups at a school for students with severe, profound and complex learning and physical disabilities in Nottingham, UK, these focus groups highlighted additional challenges that must be addressed before mainstream adoption. The **cost** of the devices was a key concern as they must be affordable for the school or individuals to consider purchasing which can be addressed by using cheap off the shelf sensors and microcontrollers. **Durability** was another concern raised as children often drop and break technological devices. To address this concern tangible interfaces specifically designed for children such toys have been developed, these provide additional protection for the electronic components but their durability remains a challenge. Students within the school also face fine and gross motor control issues making many technological solutions such as mobile apps challenging to use but by designing tangible interfaces to be easy to grip and durable it should improve the accessibility of such technology.

**Co-designing** the tangible interfaces is also vital to ensure acceptance and high usability. Designing with people with cognitive impairments is vital because they are often overlooked as "if a mental health problem presents … it is more likely to be attributed to their learning disability (diagnostic overshadowing) or classed as challenging behaviour" [8]. Co-designing with people with cognitive impairments does present additional challenges such as legality issues regarding informed consent and the impact of participation as some participants may find it challenging when confronted by their own limitations. However, many of these challenges can be overcome using method stories [10] which describe how methods work in reality instead of how they ought to work in theory.

Many of the user-oriented challenges are more difficult to overcome than the technological challenges but user adoption and engagement should be prioritised through co-design workshops to ensure the feasibility and acceptance of mental well-being tangible interfaces.

**CONCLUSION**

The emergence of mental well-being tangible interfaces and advancements in sensors, machine learning and microcontrollers suggest a new era of technological mental well-being tools, yet many challenges remain. This paper has pointed to a number of research challenges that warrant further investigation.

Many of these challenges are technological including the physical size of many off-the-shelf sensors and microcontrollers, the battery life of devices and the processing power required for on-device machine learning but with current advancements these challenges should become less problematic.

There are also many user-orientated challenges ranging from difficulties engaging users, co-design challenges and ensuring data privacy. While these challenges remain, it is imperative to involve end users at all stages of the design and development of mental well-being tangible interfaces to ensure effective interfaces are developed.

Overall mental well-being tangible interfaces present a great opportunity for individuals to automatically monitor and improve their well-being but many challenges must be overcome by researchers to make this a reality.